\documentclass[amsmath,amssymb,floatfix,a4paper]{spie}  

\usepackage{graphicx,float}
\usepackage{subfigure}
\usepackage{dcolumn}
\usepackage{bm}
\usepackage{mathrsfs}
\usepackage{txfonts}
\usepackage{CJK}
\usepackage[amssymb]{SIunits}
\usepackage{epsfig}
\usepackage{lipsum}
\usepackage{color}
\usepackage{textcomp}
\usepackage[table]{xcolor}
\usepackage{multirow}
\usepackage[T1]{fontenc}
\usepackage{times}
\usepackage{placeins}
\usepackage{inputenc}
\usepackage{placeins}
\usepackage[sort,numbers,compress]{natbib}

\title{Plasmonic nano-resonator enhanced one-photon luminescence from single gold nanorods}

\author{Keyu Xia\supit{*a,b}, Yingbo He\supit{b}, Hongming Shen\supit{b}, Yuqing Cheng\supit{b}, Qihuang Gong\supit{b,c}, Guowei Lu\supit{\dag,b,c}
\skiplinehalf
\supit{a}ARC Centre for Engineered Quantum Systems, Department of Physics and Astronomy, Macquarie University, NSW 2109, Australia; \\
\supit{b}State Key Laboratory for Mesoscopic Physics, Department of Physics, Peking University, Beijing 100871, China; \\
\supit{c}Collaborative Innovation Center of Quantum  Matter, Beijing 100871, China
}

\authorinfo{\flushleft *xiakeyu@gmail.com; Phone: 61 2 9850 8933\\ $\dag$guowei.lu@pku.edu.cn}
\begin{document}
\maketitle

\begin{abstract}
  Strong Stokes and anti-Stokes one-photon luminescence from single gold nanorods is measured in experiments. It is found that the intensity and polarization of the Stokes and anti-Stokes emissions are in strong correlation. Our experimental observation discovered a coherent process in light emission from single gold nanorods. We present a theoretical mode, based on the concept of cavity resonance, for consistently understanding both Stokes and anti-Stokes photoluminescence. Our theory is in good agreement of all our measurements.
\end{abstract}


\nopagebreak


\section{INTRODUCTION}
Electrons collective oscillations of metallic nanostructures, i.e. localized surface plasmon resonances (LSPRs), provide a powerful way for the control and manipulation of photons at the nanoscale dimension. This metallic nanostructure can be modeled as a plasmonic nano-resonator in our theoretical model. The unique optical properties of plasmonic nanostructures, such as light scattering, absorption, and emission etc., have attracted much attention. Recently, light emission from plasmonic nanoparticles has received considerable attention as metallic nanostructures exhibit remarkable optical and physical properties different from those of their bulk counterparts \cite{NL12p4385,SizeDependence1,PulsePL1,SR4p3867,PL}.
The photoluminescence (PL) of metallic nanostructure promises important applications in nanoimaging \cite{LightEmission1,PSaturate1,Guowei1,Guowei2}, nanosensing \cite{thermal2,sensing1} and even hybrid quantum systems \cite{QPAppl}. For instance, two-photon luminescence (TPL) from gold nanoparticles has been applied recently in nanosensing \cite{thermal2} and nano imaging \cite{LightEmission2} etc. Actually, the light emission from plasmonic nanostructures can be excited by electrons (e.g. cathodoluminescence) or photons (e.g. one-photon or two-photon luminescence). For one-photon luminescence (OPL) from metallic nanostructures excited by a CW laser, the Stokes emission band of nanostructures often resembles their scattering spectral line shape, which implies the strong correlation between the PL and the surface plasmon resonances. In contrast to the Stokes emission in the OPL, the anti-Stokes emission has a considerably different spectral profile but rarely receives attention, although it has been observed in the OPL of gold nanoparticles excited by a cw laser a decade ago \cite{PulsePL4}. The anti-Stokes emission has received considerable attention in two-photon luminescence (TPL) excited by a pulse laser. It is normally companied with an intensity Stokes emission band. However, previous theories for two-photon absorption process cannot explain the anti-Stokes and Stokes emission unifiedly.

Up to now, the PL from metallic nanostructures has been studied widely, nevertheless the physical origin of light emission remains a subject of debate \cite{SizeDependence1,PulsePL1,ChemPnt,PRB72p235405}. For instance, the inertialess frequency up-conversion has been reported very early by heating the metal into thousands of Kelvin using near and far IR lasers. This anti-Stokes emission was considered as a result of the thermalization dynamics \cite{thermal1,thermal3,PulsePL2}. However, the incoherent emission spectrum from this black body radiation has a Planck distribution and must be unpolarized. In contrast, the anti-Stokes OPL from single gold nanorod observed in our experiment is polarized and has a strong dependence on the excitation and detector polarizations \cite{OurExperiment}. The strong polarization-dependence of PL indicates a coherent mechanics in light emission. Very recently, the thermal occupation of electron-hole excitations in Raman scattering has been suggested to understand the anti-Stokes emission \cite{thermal2}. However, the theory is difficult to consistently explain both the anti-Stokes and Stokes components of the OPL from the gold nanorods excited by a cw laser (see Fig. 2(A) in \cite{thermal2}). The discrepancy between the prediction of the theory and the experimental measurement can be one order in amplitude for the Stokes emission and even for the anti-Stokes emission in the vicinity of the excitation frequency. Therefore, to understand the anti-Stokes PL emission requires further investigation.

In this work, we experimentally studied the OPL from a single gold nanorod excited by a CW laser, and propose another theoretical model to understand light emission process. In our experiment the nanorods was obtained by wet chemical growth method with low surface defect loss. It is found that the light emission from single gold nanorods is a coherent process. As an evidence, the intensity and polarization of both Stokes and anti-Stokes emissions are in strong correlation and they are related with the nanorod longitudinal LSPR.

Based on the observed features of the OPL from single gold nanorod, we model the OPL from single gold nanorod as \emph{coherent} emission of resonant radiation from a plasmonic nano-resonator. The key idea of our model is that the electron oscillation in plasmonic nano-resonator can enhance re-radiation of received energy through its intrinsic LSPR mode. In addition, our model reveals that the free electrons distribution at thermal equilibrium would also affect the OPL spectral profile. Therefore, both the plasmonic resonant enhancement and the free electrons distribution determinate the PL spectral shape. Our model also provides a good explanation for the main features of previous experimental observations, e.g. the enhanced quantum yield (QY) and the shape dependence of the QY of OPL. This model provides a self-consistent and unified understanding for both the anti-Stokes and Stokes OPL emission properties. Moreover, TPL's time dynamics has been measured and explained phenomenologically \cite{thermal2,PulsePL4,TPL2}, but a microscopic explanation is lack. Using present model, we can further explain the origin of time dynamics and spectral feature of TPL from a single gold nanorod, and the dynamics and spectral feature of TPL. 

\section{EXPERIMENTAL METHOD}
\begin{figure}
 \centering
 \includegraphics[width=0.45\linewidth]{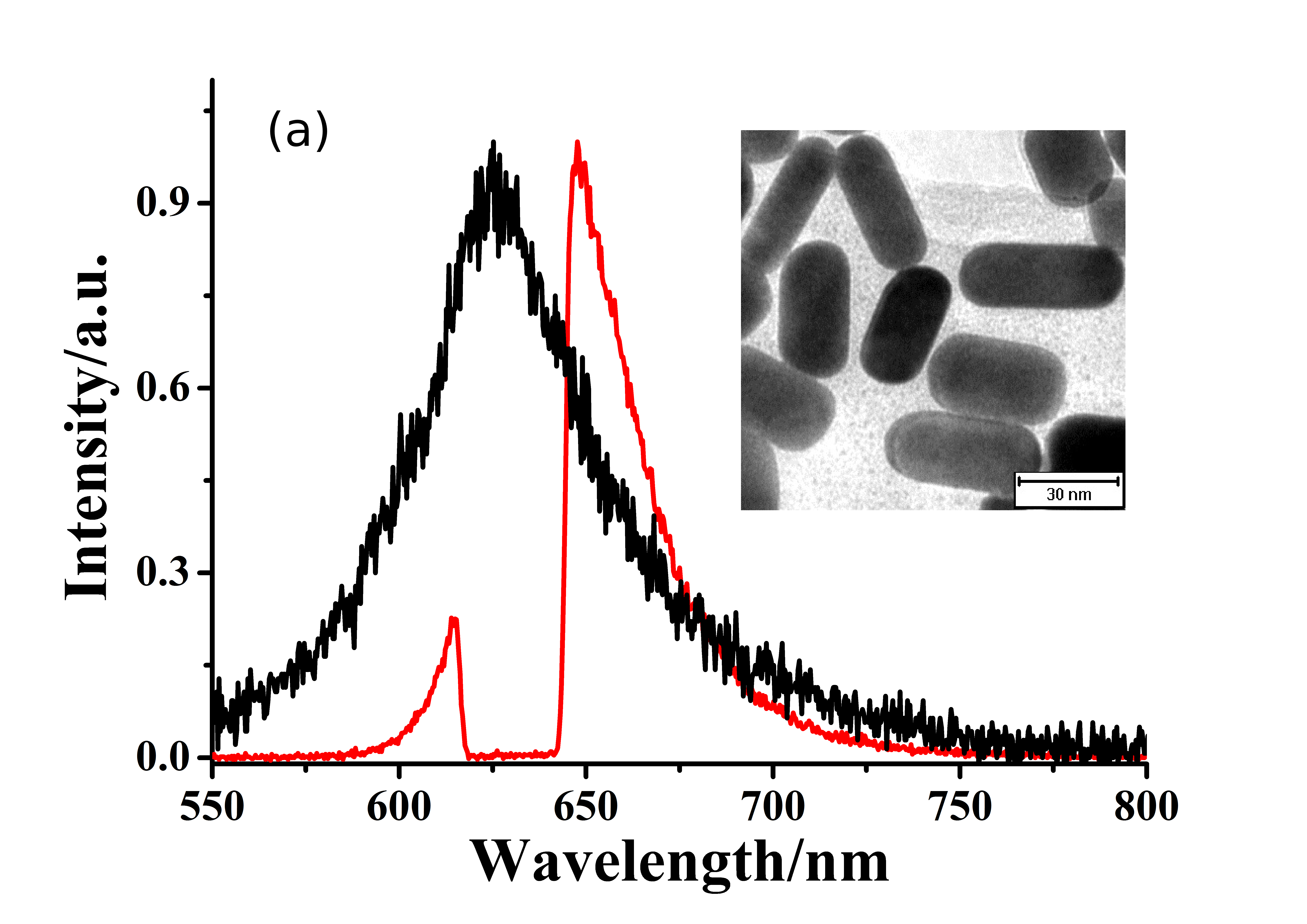}
 \includegraphics[width=0.35\linewidth]{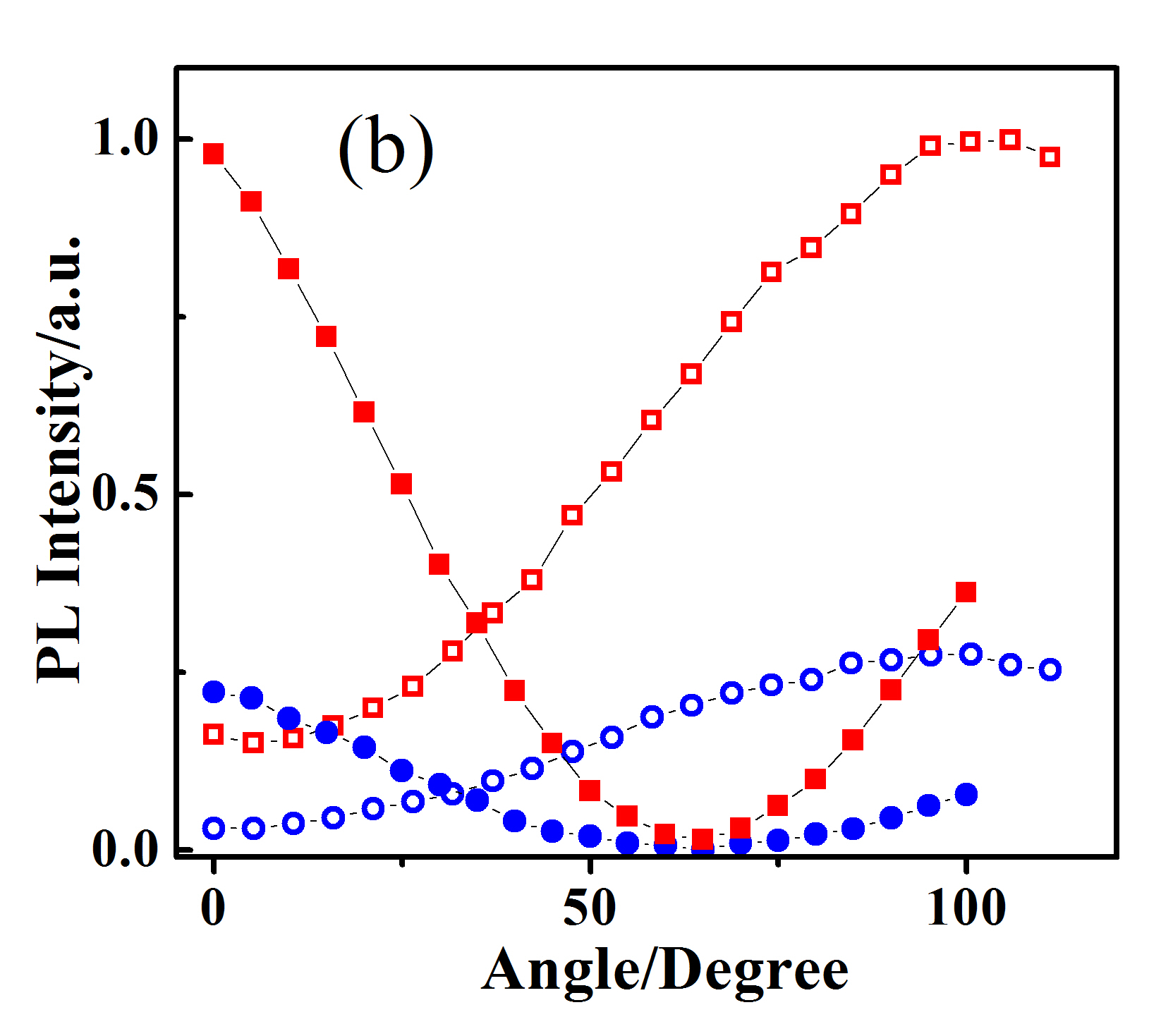}
\caption{(Color online) (a) Normalized scattering (Sc.) and PL spectra of a representative gold nanorod. The red curve is the PL spectra excited by a cw $632.8$\nano\meter~ laser. The black curve is scattering spectrum of the same nanorod. The inset presents the typical TEM image of the synthesized nanorods. (b) The Stokes (red) and anti-Stokes (blue) emission intensity present a similar polarization dependent behaviors
as excitation polarization (filled) and collection polarization (hollow)
angles being tuned.}\label{fig1}
\end{figure}

The gold nanorods investigated here were synthesized chemically, and the inset of Fig. \ref{fig1} shows a representative TEM image. The nanorods have short axes of 20-30 nm and lengths of 50-80 nm presenting different longitudinal resonant frequencies. The synthesized nanorods were immobilized onto glass coverslip with an average interparticle spacing of several micrometers for single particle spectroscopy measurements. The surface of gold nanorods is atomic smooth due to chemical growth method, and the LSPRs deviates from the interband transition resulting in a drastic reduction of the plasmon dephasing rate \cite{PRL88p077402}. These factors allow us to assume single nanorod as plasmonic resonator because of its low loss and the small affect of LSPR mode coupling. The optical measurement system (NTEGRA Spectra, NT-MDT) is based on an inverted optical microscope  which allows us in {\it situ} to obtain the scattering and PL spectra of the same nanoparticle \cite{JPCC116p25509,AOM1p335,SR4p3867}. A cw He-Ne laser at wavelength of $632.8$ \nano\meter~ ($1.96$ \electronvolt~ photon energy) was applied to excite {\it sp}-band electrons in the gold due to low photon energy avoiding excitation of interband transitions. Here, only the longitudinal plasmon model is mainly related with light emission. The excitation laser of $632.8~\nano\meter$ should dominantly excite the sp-band electrons of gold material. All these factors are benefit to reduce the complexity of the light emission system.  Meanwhile Notch filter (NF03-633E, Semrock) was utilized to obtain the Stokes and anti-Stokes emission simultaneously. The scattering spectra were recorded in {\it situ} with dark field white light total internal reflection technique, while the PL spectra of the same nanorod were excited by a cw laser at wavelength of $632.8$ \nano\meter. All spectra were measured at room temperature. More details are presented in \cite{OurExperiment}. The PL spectrum from one sample is shown in Fig. \ref{fig1}.

Interestingly, an obvious anti-Stokes emission accompanies
always the conventional Stokes emission when the nanorods
are excited by $632.8 \nano\meter$ laser. The intensity of anti-Stokes emission is even comparable to the Stokes, see Fig. \ref{fig1}(a). Anti-Stokes photoluminescence from single noble metal nanoparticles has been previously observed and are understood as radiation of high-temperature hot electrons \cite{HotE,SERSBackgrounds,LightEmission1}. However, the polarization characteristic of anti-Stokes photoluminescence attracts rarely attention. We compare the polarizations of both anti-Stokes and Stokes emission from single gold nanorods in Fig. \ref{fig1}(b). It is found that the excitation polarization, and the collection polarization dependence of the Stokes and anti- Stokes emission are strongly correlated. This experimental observation provides a proof that both emissions must result from a common process.

\section{MODEL}
\begin{figure}[h]
 \centering
\includegraphics[width=0.5\linewidth]{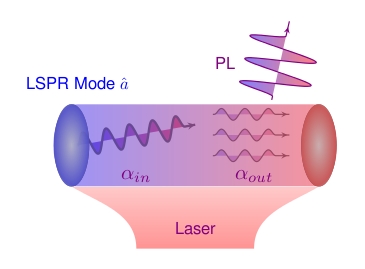} 
\caption{(Color online) Diagram showing the photoluminescence from a single gold nanorod excited by a laser beam. The laser beam drives a collective oscillation of free electrons (COFE) polarized along the polarization of the light. This electron oscillation as an input $\alpha_{in}$ subsequently excites the LSPR mode. In return, the LSPR mode generates COFE as an output $\alpha_{out}$ and emit the photon-luminescence.}\label{fig:model}
\end{figure}

We first discuss the OPL process and then we will study the time dynamics and spectrum of the TPL. In noble metallic nanostructures like gold nanorods, the sp-band free electrons can interact with photons directly and efficiently, i.e. the free electrons can be driven to collectively oscillate by an external electromagnetic field \cite{MapSPR1,MapSPR2}. Such electron oscillation would form a collective oscillation of free electrons (COFE) in metallic nanostructures when the nanostructures supported LSPR mode matches with external electromagnetic field \cite{MapSPR1,MapSPR2}. Once the LSPR mode is excited, it will decay and radiate photons to free space. Note that in order to simplify the problem, the interband transition of gold material has been ruled out in present model. On the basis of the concept of the COFE, we present a theoretical model to explain the main features of the OPL spectrum from a single gold nanorod modeled as a plasmonic nano-resonator under a laser excitation. Our model shown by the diagram in Figure \ref{fig:model} involves four processes: (i) the electromagnetic field drives the free electrons to oscillate; (ii) the oscillated electrons coupling with LSPR mode $\hat{a}_c$ as the input $\alpha_\text{in}$ ; (iii) the LSPR mode generates an output αout surrounding the surface of the nanorod; (iv) the LSPR mode-generated radiation couples to the free space and generates the OPL.

A linear-polarized input electromagnetic field $\vec{E}_\text{in}e^{-i\omega_{in}t}$ drives the free electrons to collectively oscillate at the frequency $\omega_{in}$. The polarization of the electron oscillation created by the light source is along $\vec{E}_\text{in}$. Similar to the nanoscale photonic crystal cavity localizing a quantized electromagnetic field, the single metal nanorod supports a LSPR mode $\hat{a}_c$ polarized along $\vec{d}$. The LSPR mode can be excited by the free electrons oscillation in the nanorod. That is to say, the collectively oscillating free-electron gas as an input $\alpha_\text{in}$ drives the LSPR mode polarized along either the longitudinal axis or the transversal axis of the nanorod. Here we focus on the longitudinal mode with a polarization direction $\vec{d}$. In return, the LSPR mode results in a resonant COFE with a polarization of $\vec{d}$ as an output $\alpha_\text{out}$. As a result, the oscillating free-electron gas emits photons as OPL. In our model the free electrons gain energy directly from photons but not from the heating effect of the excitation laser as claimed in \cite{thermal2,thermal1,thermal3,PulsePL2}.

The coupling efficiency between the light and the COFE is proportional with the spatial and temporal overlap of the modes \cite{Couplingeff1,Couplingeff2}, 
\begin{equation} \label{eq:etac}
 \eta_\text{c}(\omega)  = \frac{\langle H_\text{COFE} | H_\text{photon}\rangle^2}{|\langle H_\text{COFE}\rangle|^2 |\langle H_\text{photon}\rangle|^2} \;,
\end{equation}
where $H_\text{COFE}$ and $H_\text{photon}$ are the profiles of the COFE and incident light modes. $H_\text{COFE}$ of the LSPR mode varies with different sizes and aspect ratios. Moreover, the coupling efficiency is also determined by the occupation of electrons, $\rho_\text{e}(\omega)=|\langle H_\text{COFE}\rangle|^2$, at the energy $\hbar \omega$. Thus the coupling efficiency is given by
\begin{equation} \label{eq:coupling}
 \eta_\text{in/out}(\omega)  = \eta_\text{c}(\omega) \rho_\text{e}(\omega)\,,
\end{equation}
for the excitation process (input) and the emission process (output).
 The occupation $\rho_\text{e}(\omega)$ is proportional to the total number of free electrons, $N_\text{FE}$, attending the collective oscillation and corresponds to the radiation power at the frequency $\omega$. Our definition is different from those in \cite{Couplingeff1, Couplingeff2} because we consider the dependence of electron occupation, $\rho_\text{e}(\omega)$, on the energy $\hbar\omega$ of the sp-band free electrons. This occupation in Eq. (\ref{eq:coupling}) is crucial for understanding the anti-Stokes emission. 

Now we turn to derive the OPL spectrum from the Hamiltonian describing the interaction between the light and the plasmon mode $\hat{a}_\text{c}$. When we apply a cw laser beam to the gold nanorod, the sp-band free-electron gas first gains the photo energy $\hbar \omega_\text{in}$ and starts to oscillate collectively. Thus we have $\rho_\text{e}(\omega) = N_\text{FE}$ in the input coupling, $\eta_\text{in}$. The oscillating free electrons decay rapidly due to the fast process of plasmon decay. The electrons and lattice reach a thermal equilibrium due to the fast electron-phonon interaction \cite{ESD1,FermiDiracD}. Therefore, the output coupling, $\eta_\text{out}$, is also dependent on the free-electron state density (FESD) $\rho_\text{e}(\omega)$ at the energy $\hbar\omega$. In the thermal equilibrium, the FESD follows the Fermi-Dirac statistic distribution \cite{ESD1,FermiDiracD}
\begin{equation}
 \rho_\text{e}(\omega) = N_\text{FE}/(1+ e^{\hbar (\omega - \mu)/K_B T}) \,,
\end{equation}
with the so-called chemical potential. $K_\text{B}$ is the Boltzmann constant and $T$ is the localized thermal equilibrium temperature. Note that this excitation distribution is essentially different from the thermal occupation suggested by Huang et al. \cite{thermal2}. Although this process approaching thermal equilibrium is incoherent, the absorption of excite laser and PL are coherent. This is the key  point different from Huang's explanation \cite{thermal2}. 

The motion of the SPR mode can be described using the Hamiltonian \cite{Driving,xia1}
\begin{equation}\label{eq:H}
 H=\omega_\text{c} \hat{a}_\text{c}^\dag \hat{a}_\text{c} + i\sqrt{2\kappa_{ex1}} \left(\alpha_\text{in}^* \hat{a}_\text{c}^\dag  - \alpha_\text{in}  \hat{a}_\text{c}\right) \,,
\end{equation}
where the first term describes the free-energy of the LSPR mode $\hat{a}_\text{c}$ and $\omega_\text{c}$ is the resonance frequency. The COFE relevant to $\alpha_\text{in}$ drives the LSPR mode with a coupling rate $\kappa_\text{ex1}$. Thus, the input $\alpha_\text{in}$ is determined by the amplitude of the COFE and the angle $\theta$ defined by $\cos\theta=\vec{E}\cdot \vec{d}/|\vec{E}| |\vec{d}|$ and takes the form 
\begin{equation} \label{eq:input}
 \alpha_\text{in} = \sqrt{\eta_\text{in}(\omega_\text{in})} cos\theta E_\text{in}e^{-i\omega_\text{in}t} \,.
\end{equation}
Here we replace the driving by a classical number $\alpha_\text{in}$ as the quantum average of laser field. It is reasonable when the driving laser is strong and includes many photons \cite{Driving,xia1}.

The dynamics of the LSPR mode can be solved by the equation \cite{xia1,xia2,xia3}
\begin{equation} \label{eq:OPLt}
 \frac{\partial \hat{a}_\text{c}}{\partial t} = -(i\omega_\text{c} + \kappa) \hat{a}_\text{c} + \sqrt{2\kappa_\text{ex1}}\sqrt{\eta_\text{in}(\omega_\text{in})}\cos\theta E_\text{in}e^{-i\omega_\text{in}t} \,,
\end{equation}
where $\kappa$ is the total decay rate of the LSPR mode due to its coupling to the COFE, the environment and the material absorption.
The time-dependent $\hat{a}_\text{c}(t)$ is solved to be
\begin{equation}\label{eq:evolution}
 \hat{a}_\text{c}(t) =  \hat{a}_\text{c}(0) e^{-(i\omega_\text{c} + \kappa)t} - \sqrt{2\kappa_\text{ex1} \eta_\text{in}(\omega_\text{in})}\cos\theta \frac{E_\text{in}^* e^{-i\omega_\text{in}t}}{i\Delta_\text{c} + \kappa} 
  + \sqrt{2\kappa_\text{ex1} \eta_\text{in}(\omega_\text{in})}\cos\theta \frac{E_\text{in}^* e^{-(i\omega_\text{c} + \kappa)t}}{i\Delta_\text{c} + \kappa} \,.
\end{equation}
The excitation of the LSPR mode includes three contributions. The first term indicates the oscillation due to the initial occupation of the LSPR mode itself. It corresponds to the luminescence spectra excited by a ultrashort laser pulse and has the same spectral profile as the scattering featured by the LSPR mode \cite{PulsePL1,PulsePL4, PRB68p115433}. It is absent in the OPL here. The second term origins from the elastic scattering of the input field. We assume that the incident light has a linewidth of $\gamma_\text{in}$ spanning a spectrum over $\delta\lambda_\text{in}$ that the power spectrum of the input $\Re\left[\int_0^\infty \lim_{t \to \infty} \langle E_\text{in}^*(t+\tau)E_\text{in}(t)\rangle e^{-i\omega \tau}d\tau\right]=S_\text{in}(\omega)\propto I_\text{in}$, where $I_\text{in}= |E_\text{in}|^2$ is the intensity of the input laser field. $\Re[]$ means the real part of numbers. Because $\gamma_\text{in}$ (corresponding to $\delta\lambda_\text{in}\sim 0.1 ~\nano\meter$ in our experiment) of the input laser field is much smaller  than the decay rate $\kappa$ of the LSPR modes, this term is much larger than the third contribution in the intensity of the PL spectra. But it is usually filtered before entering the detector when measuring  the PL emission. On the other hand, this term also responds to the scattering spectra excited by an incoherence white light and has Lorentzian profiles, $\propto \frac{\kappa^2}{(\omega - \omega_\text{c})^2 + \kappa^2}$. We are interested in the third term, which corresponds to the PL emission of interest. It includes both the Stokes and anti-Stokes components in the PL spectra.

Using the input-output relation $\langle \hat{a}_\text{out} \rangle = \sqrt{2\kappa_\text{ex2}} \langle \hat{a}_\text{c} \rangle$ \cite{InputOutput1,InputOutput2}, where $\langle \hat{Q} \rangle $ estimates the quantum average of operator $Q$ and $\kappa_\text{ex2}$ the outgoing coupling, and taking into account the state density of free electrons $\rho_\text{e}(\omega)$, the optical field emitted by the nanorod takes the form
\begin{equation}\label{eq:Output}
 \hat{E}^\dag_\text{out}(t) = \sqrt{\eta_\text{out}(\omega)} \hat{a}^\dag_\text{out}  \,.
\end{equation}
The detected intensity of light emission from the single gold nanorod can be evaluated by
\begin{equation}\nonumber
 I_{full}(\omega) =  \frac{\eta_\text{D}(\omega)\cos^2\beta}{\pi} \Re\left[ \int_{0}^{\infty} \lim_{t \to \infty}\langle E^\dag_\text{out}(\tau+t)E_\text{out}(t)  \rangle e^{-i\omega \tau} d\tau\right] \,
\end{equation}
where $\eta_\text{D}(\omega)$ is the quantum efficiency of the detector and $\beta$ the detection polarization angle between the detector and the polarization of the emitted light. The emission intensity $I_\text{full}(\omega)$ includes two contributions: the PL $I_\text{PL}(\omega)$ and the scattering $I_\text{SC}(\omega)$.

According to the second term of Eq. (\ref{eq:evolution}), the scattering $I_\text{SC}$, which can be obtained experimentally through dark-filed white light scattering, reads
\begin{equation}\label{eq:SComega}
 I_\text{SC}(\omega) = \frac{4\langle\cos^2\beta\rangle \langle\cos^2\theta\rangle\eta_\text{D}(\omega)\eta_\text{in}(\omega_\text{in})}{\pi}  \eta_\text{out}(\omega)
  \frac{ \kappa_\text{ex1}\kappa_\text{ex2} }{\Delta_\text{c}^2  + \kappa^2} \delta(\omega-\omega_\text{in}) \otimes S_\text{in}(\omega)\,,
\end{equation}
where $\delta(\omega)$ is a delta function of $\omega$ and $\otimes$ means the convolution of two functions. $\langle \cdot \rangle$ evaluates the statistic average value. The power spectrum of the white light is approximately constant. Thus $I_\text{SC}(\omega)$ has the Lorentzian profile. Note that $\eta_\text{in}(\omega)\delta(\omega-\omega_\text{in})=\eta_\text{in}(\omega_\text{in})\delta(\omega-\omega_\text{in})$.

Different from the scattering, the OPL is excited by a narrow band laser pulse. After filtering the input laser field and using the quantum regression theorem \cite{QR}, we obtain the OPL intensity readout by the photon detector as
\begin{equation}\label{eq:PLomega}
 I_\text{PL}(\omega) = \frac{4\cos^2\beta\cos^2\theta\eta_\text{D}(\omega)\eta_\text{in}(\omega_\text{in})}{\pi}  \eta_\text{out}(\omega)
  \frac{ \kappa_\text{ex1}\kappa_\text{ex2} }{\Delta_\text{c}^2  + \kappa^2} \frac{\kappa}{(\omega - \omega_\text{c})^2  + \kappa^2} \otimes S_\text{in}(\omega)\,.
\end{equation}
Because the linewidth $\gamma_\text{in}$ of the input laser is much smaller than $\kappa$, to a good approximation, we have $S_\text{in}=I_\text{in}\delta(\omega)$ so that $\frac{\kappa}{(\omega - \omega_\text{c})^2  + \kappa^2} \otimes S_\text{in}(\omega)= I_\text{in}\frac{\kappa}{(\omega - \omega_\text{c})^2  + \kappa^2}$. As a result, the PL spectrum $I_\text{PL}(\omega)$ is a Lorentzian profile modulated by the occupation of free electrons $\rho_\text{e}(\omega)$.

To compare our model with the experiment results, we rewrite the scattering and PL spectra as a function of the wavelength. The scattering spectra has a Lorentzian profile and simply takes the form
\begin{equation}\label{eq:SC}
 I_\text{SC}(\lambda) = A' I_\text{in}\eta_\text{D}(\lambda)  \frac{\delta\lambda_\text{c}^2}{(\lambda_\text{c}^2/\lambda - \lambda_\text{c})^2  + \delta\lambda_\text{c}^2} \,,
\end{equation}
where $A'=\mathcal{C}\frac{\eta_\text{in}^2(\omega_\text{in})\kappa_\text{ex1}\kappa_\text{ex2}}{\kappa^2\pi}$ is a constant, $\langle\cos^2\beta\rangle=1/2$ and $\langle\cos^2\theta\rangle=1/2$ for unpolarized excitation, $\lambda_\text{c}=\frac{2\pi c}{\omega_\text{c}}$ is the light wavelength corresponding to the resonance frequency $\omega_\text{c}$ of the plasmon resonator with the light velocity $c$ and $\delta\lambda_\text{c}=\frac{\kappa}{\omega_\text{c}}\lambda_\text{c}$ the linewidth of the radiation due to the LSPR mode in \nano\meter. Thus, $I_\text{SC}(\lambda)$ is a Lorentzian profile modulated by the quantum efficiency $\eta_\text{D}(\omega)$.

In contrast to the scattering spectrum, the PL spectrum is dependent on the occupation of free electrons $\rho_\text{e}(\omega)$ and has the form
\begin{equation}\label{eq:PLLambda}
 I_\text{PL}(\lambda) =  AI_\text{in}\cos^2\beta\cos^2\theta \eta_\text{in}(\lambda_\text{in})\eta_D(\lambda)\rho_\text{e}(\lambda)  
 \frac{ \delta\lambda_\text{c}^2 }{\left( \lambda_\text{c}^2/\lambda_\text{in} - \lambda_\text{c}\right)^2  + \delta\lambda_\text{c}^2} 
\frac{\delta\lambda_\text{c}^2}{(\lambda_\text{c}^2/\lambda - \lambda_\text{c})^2  + \delta\lambda_\text{c}^2}\,,
\end{equation}
where $A=\frac{4\kappa_\text{ex1}\kappa_\text{ex2}}{\kappa^3\pi} \eta_\text{c}(\lambda)$ is a constant. $\lambda_\text{in}=\frac{2\pi c}{\omega_\text{in}}$ is the wavelength of the input cw laser field. Here the FESD is rewritten as 
\begin{equation} \label{eq:FESD}
 \rho_\text{e}(\lambda)= N_\text{FE}/ (1+e^{\tau_T (\lambda_\mu/\lambda -1)}) \,,
\end{equation}
where $\lambda_\mu=2\pi c/\mu$ is the corresponding light wavelength of the chemical potential and $\tau_T=\hbar\mu/K_\text{B}T$ is a transition energy related to environment temperature $T$. Because the electrons gain the energy of $\hbar \omega_\text{in}$, we have the chemical potential $\mu\approx \omega_\text{in}$ yielding $\lambda_\mu\approx\lambda_\text{in}$. Our experimental observations and the reported results by others \cite{PulsePL2,ChemPnt} support this assumption. According to Eq. (\ref{eq:FESD}), we have $\rho_\text{e}(\lambda_\text{in})=0.5N_\text{FE}$. Later we will study the dependence of the parameter $\tau_T$ on the input power.

Interestingly, The formula Eq.~(\ref{eq:PLLambda}) immediately provides an explanation of the aspect-ratio dependence of PL emission discussed by Mohamed et al. \cite{SizeDependence1,SizeDependence2}. According to our model and Eq.~(\ref{eq:PLLambda}), similar to an optical cavity, the peak position $\lambda_\text{c}$ (resonance wavelength) of PL emission increase linearly as the nanorod length $L$ increasing. Note that this effective length $L$ corresponds to the mode volume of the one-dimensional plasmonic nano-resonator.  $\eta_\text{c}(\omega)$ is a linear function of $|H_\text{COFE}|^2$ which is proportional to the number $N_\text{FE}$ of free electrons in nanorods. In a gold nanorold, it is reasonable to assume that $N_\text{FE}$ is proportional to the length of nanorod. Thus, the quantum efficiency corresponding to $I_\text{OPL}$ increase quadratically as a function of the length because $\eta_\text{in}(\omega_\text{in})\eta_\text{out}(\omega) \propto N^2_\text{FE}$. These predictions agree well with experiments \cite{SizeDependence1,SizeDependence2,NL12p4385} and Mohamed's model as well \cite{SizeDependence1,SizeDependence2}. 
As the length increases further, the active number of free electrons becomes saturate due to the temporal and spatial correlation limited by the free path effect of electrons \cite{MFEP}.
Our model taking the concept of the COFE correctly predicts that the OPL emission becomes saturate as the length of nanorod increases to about $50~ \nano\meter$ \cite{SizeDependence1,SizeDependence2,NL12p4385}. 
On the other hand, the OPL can be saturate as increasing the input power when the excitation laser actives all free electrons. This kind of saturation has been observed in others' experiments \cite{PSaturate1,thermal2} and our observation as well \cite{OurExperiment}.

According to our model, the detected OPL has the features as follows: (i) the OPL intensity is a linear function of the intensity of the input laser field; (ii) the OPL intensity depends on both of the polarization of the input laser field and that of the LSPR mode. It is modulated by the function of $\cos^2(\theta)$; (iii) the PL intensity is dependent on the detuning between the driving laser field and the LSPR mode; (iv) the OPL is Lorentzian profile modulated by the quantum efficiency $\eta_\text{D}$, the FESD $\rho_\text{e}$; (v) the OPL emission becomes saturate when the input power excites all $N_\text{FE}$ free electrons to a collective oscillation \cite{PSaturate1,OurExperiment,thermal2}; (vi) the OPL efficiency is dependent on the aspect ratio of the nanorods.

Our model can also explain the enhanced quantum yield (QY) of gold nanoparticles rather than the bulk materials.
To estimate the QY we assume a constant decay rate $\kappa$ and the external coupling rates $\kappa_\text{ex1}$ and $\kappa_\text{ex2}$ for simplicity although the decay rate is dependent on the length of nanorods \cite{DecaySize1, DecaySize2}. We consider the integral spectrum from a low enough boundary $\omega_{cut}\approx 1.424$  \electronvolt~ corresponding to $869$ \nano\meter~ to $\omega_\text{in}$ for the Stokes OPL emission and from $\omega_\text{in}$ to $+\infty$ for the anti-Stokes OPL. The QY can be evaluated as
\begin{equation}\label{eq:QYS}
  QY_\text{S}(\omega_\text{in},L)  =  \frac{BL^2\kappa^2}{(\omega_\text{in}-\omega_\text{c}(L))^2 + \kappa^2} \int_{\omega_\text{cut}}^{\omega_\text{in}} \frac {\omega_\text{in}}{\omega}\frac{\kappa \eta_\text{out}(\omega,\omega_\text{in})}{(\omega-\omega_\text{c}(L))^2 + \kappa^2} d\omega \,,
\end{equation}
for the Stokes OPL, and 
\begin{equation}\label{eq:QYAS}
   QY_\text{AS}(\omega_\text{in},L)  =  \frac{BL^2\kappa^2}{(\omega_\text{in}-\omega_\text{c}(L))^2 + \kappa^2}  \int_{\omega_\text{in}}^{+\infty}\frac{\omega_\text{in}}{\omega} \frac{\kappa \eta_\text{out}(\omega,\omega_\text{in})}{(\omega-\omega_\text{c}(L))^2 + \kappa^2} d\omega \,,
\end{equation}
for the anti-Stokes OPL. $B$ is a scaling coefficient. In $\eta_\text{c}(\omega)$ the chemical potential $\mu$ is replaced by $\omega_{in}$ for simplicity.  The value of $\omega_\text{cut}$ is not crucial and it can be chosen smaller or larger. For a bulk film the total number of the active free electrons $N_\text{FE,Bulk}$ is constant because the free electrons have a limited correlation length. To provide a rough estimation, we assume that $N_\text{FE,Bulk}$ corresponds to a length $L_\text{max}=50$ \nano\meter. Then the QY of the OPL from a bulk material is roughly given by
\begin{equation}\label{eq:QYBulk}
  QY_\text{Bulk}(\omega_\text{in},L)  = \frac{BL_\text{max}^2\kappa^2}{(\omega_\text{in}-\omega_\text{c}(L))^2 + \kappa^2}  \int_{\omega_\text{cut}}^{\omega_\text{in}} \frac {\omega_\text{in}}{\omega}\frac{\kappa \eta_\text{out}(\omega,\omega_\text{in})}{(\omega-\omega_\text{c}(L))^2 + \kappa^2} d\omega \,.
\end{equation} 
It can be roughly evaluated as $QY_\text{Bulk}(\omega_\text{in},L\to \infty)$. In following investigation, we will compare QYs from single gold nanorods and bulk material.

So far, we investigate the features and explain the origin of OPL. Next we will discuss the time dynamics and spectrum of TPL from a single gold nanorod excited by a ultrashort laser pulse, $\vec{E}_\text{in}(t)e^{-i\omega_\text{in}t}$. The basic mechanism of TPL from a single gold nanorod is similar to that of the OPL process. The COFE is excited after absorbing two photons from a laser pulse and subsequently drives the plasmonic resonantor mode $\hat{a}_\text{c}$. Thus the input of the plasmonic resonator is given by
\begin{equation}
 \alpha_\text{in}(t) = \sqrt{\eta_\text{TPL}}\cos^2\theta |E_\text{in}(t)|^2 e^{-2i\omega_\text{in}t} \,,
\end{equation}
where  $\eta_\text{TPL}$ is the two-photon efficiency. Normally, TPL is excited with ultrashort laser pulses, of which the duration is much shorter than the time reaching FESD thermal equilibrium due to the electron-electron and electron-phonon interaction. Thus $\eta_\text{TPL}$ is a constant in time to a good approximation. 

The dynamics of the LSPR mode under two-photon excitation is governed by the equation
\begin{equation} \label{eq:TPLt}
 \frac{\partial \hat{a}_\text{c}}{\partial t} = -(i\omega_\text{c} + \kappa) \hat{a}_\text{c} + \sqrt{2\kappa_\text{ex1}}\alpha_\text{in}(t) \,.
\end{equation}
We obtain the time evolution of the plasmonic mode that $\hat{a}_\text{c}(t) = \sqrt{2\kappa_\text{ex1}} \sqrt{\eta_\text{TPL}}\cos^2\theta e^{-i\omega_c t}\int^t_{0_{-}} |E_\text{in}(\tau)|^2  e^{i(\omega_c-2\omega_\text{in})\tau} e^{\kappa (\tau-t)}d\tau$, where $0_{-}$ is the time immediately before the excitation pulses is applied.
When the excitation pulses is short enough, $\hat{a}_\text{c}(t) \approx \hat{a}_\text{c}(0_+) e^{-(i\omega_\text{c} + \kappa)t}$ with $t>0_+$, where $\hat{a}_\text{c}(0_+)$ is the excitation immediately after the ultrashort laser pulse. $\langle \hat{a}^\dag_\text{c}(0_+)\hat{a}_\text{c}(0_+)\rangle$ is proportional to $\eta_\text{TPL}\cos^4\theta I_\text{in}^2$, where $I_\text{in} = |E_\text{in,0}|^2$ and $E_\text{in,0}$ is the amplitude of the ultrashort excitation laser. We normalize the excitation $\hat{a}_\text{c}$ such that $\langle \hat{a}^\dag_\text{c}(0_+)\hat{a}_\text{c}(0_+)\rangle = \eta_\text{TPL}\cos^4\theta I_\text{in}^2 \langle \hat{\tilde{a}}^\dag_\text{c}(0_+)\hat{\tilde{a}}_\text{c}(0_+)\rangle$.

The TPL intensity emitted by the single gold nanorod then reads 
\begin{equation}\label{eq:TPLoutput}
 I_\text{out}(\omega,\omega_\text{in})=\eta_\text{TPL}\cos^4\theta I_\text{in}^2 \langle \hat{\tilde{a}}^\dag_\text{c}(0_+)\hat{\tilde{a}}_\text{c}(0_+)\rangle \frac{\kappa}{(\omega -\omega_\text{c})^2 + \kappa^2}\,.
\end{equation}
Obviously, the TPL intensity quadratically increases as the input intensity increasing, i.e. $I_\text{out} \varpropto I^2_\text{in}$, and follows the $\cos^4\theta$ polarization dependence, as reported in experiments \cite{LightEmission1, SPIE}. The TPL spectrum resembles the scattering. This is in good agreement with previous experiments \cite{LightEmission1, PulsePL2}.

Next we test the validity of our model by fitting the scattering and OPL spectra observed in our experiments.
 
\section{RESULTS}
\subsection{One-photon luminescence}
In this section we have reproduced both the scattering spectrum and the OPL. We retrieved the quantum efficiency of the detector from the manufacturer's manual and then fitted it using a polynomial function. The quantum efficiency of the detector of interest from $500$ \nano\meter ~ to $800$ \nano\meter~ can be perfectly fitted by a polynomial function $\eta_D(\lambda)= - 8.87998\times 10^{-11} \lambda^4 + 2.3287\times 10^{-7} \lambda^3 - 2.3242 \times  10^{-4} \lambda^2 + 0.10378 \lambda - 16.421$ with $\lambda$ has the unit of \nano\meter. This quantum efficiency will be used to fit the experimental data using our model.

\begin{figure}
 \centering
 \includegraphics[width=1\linewidth]{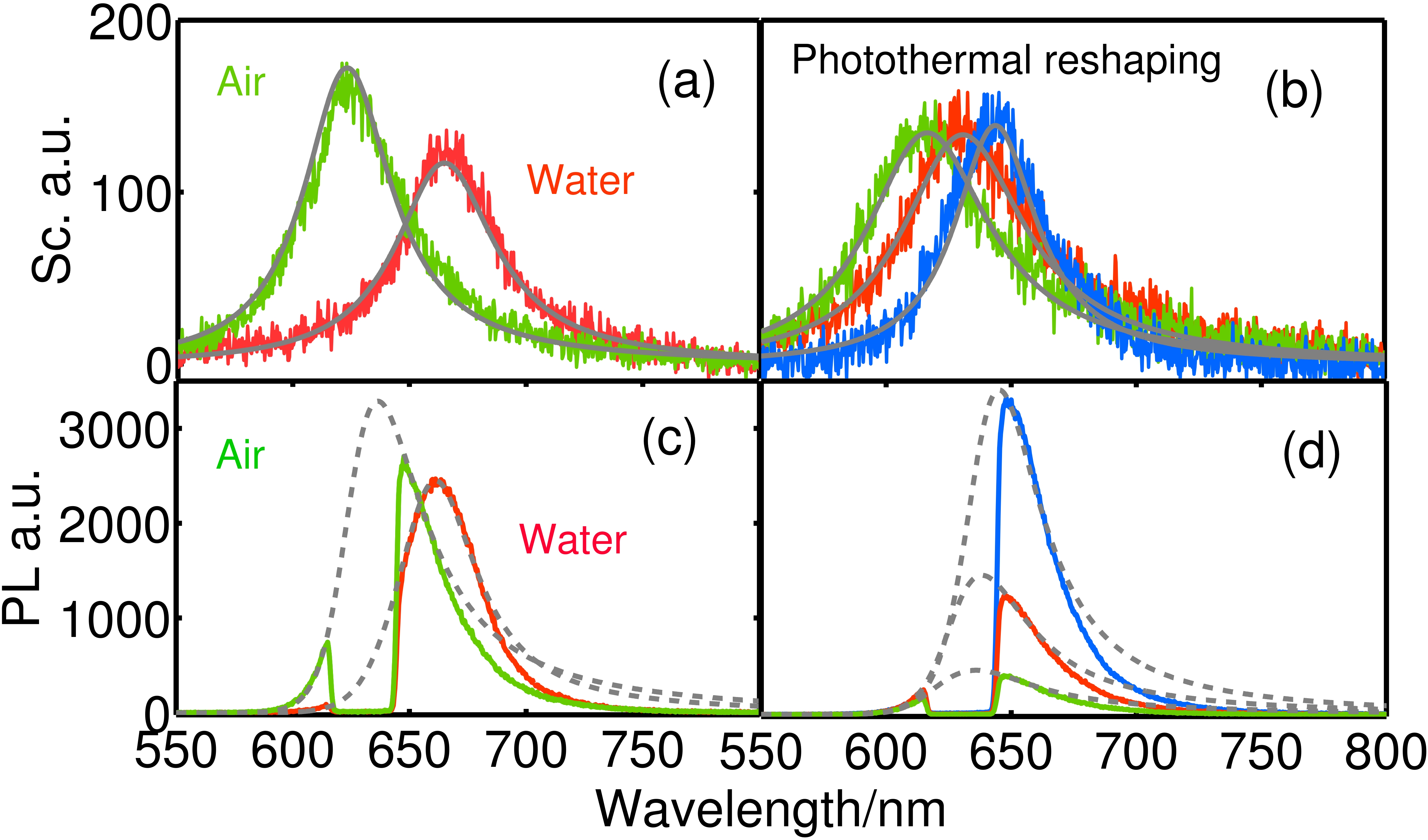}\\
\caption{(Color online) Light emission spectra as the local surface plasmon resonant frequencies being changed. (a) and (c) The scattering and PL spectra of the same nanorod by changing the refractive index in situ. (b) and (d) The scattering and PL spectra of the single individual nanoparticle being reshaped through photothermal effect. All spectra were recorded after photothermal reshaping with the same excitation intensity. To use the y-axis ticks in (c), the PL spectra in (d) is scaled by a factor of $2$. PL spectra are fitted by using Eq. (\ref{eq:PLLambda}) with Fermi-Dirac statistics.}\label{fig:PL}
\end{figure}

Note that the scattering spectra are created by the unpolarized, weak white light shedding on the gold nanorod. The white light has a very broad and flat spectra over $500$ \nano\meter~ to $800$ \nano\meter.
As predicted by Eq. (\ref{eq:SC}), the scattering spectra have Lorentzian profiles. Our formula Eq. (\ref{eq:SC}) excellently reproduces the observed scattering spectra shown in Figs. \ref{fig:PL} (a) and (b), using the parameter shown in Tab. (\ref{tbl:a}).

\begin{center}
\begin{table}[H]
\centering
\caption{Parameters for scattering spectra in Figs. \ref{fig:PL} (a) and (b)}
\begin{tabular}{l|c|c|c|c|c|c|c|r}
\hline \hline
 & Line \# & $A'$ & $I_{in}$ [counts]& $\lambda_{in}$ [\nano\meter] & $\lambda_{c}$ [\nano\meter] & $\lambda_{in} - \lambda_{c}$ [\nano\meter] & $\delta\lambda_c$ [\nano\meter] & $\kappa$ [\electronvolt]  \\ \hline
 \multirow{2}{*}{SC. in (a)} & Green & $180$ & \texttwelveudash & \texttwelveudash & 622.9 & \texttwelveudash  & $23$ &  $0.0733$ \\ \cline{2-9}
 & red & $126$  & \texttwelveudash & \texttwelveudash & $664.9$ & \texttwelveudash  & $26$ &  $0.0727$ \\ \hline \hline
 \multirow{2}{*}{SC. in (b)} & blue & 2940 &\texttwelveudash  & \texttwelveudash & 644.1 & \texttwelveudash & 20 & $0.0596$ \\ \cline{2-9}
& red & 4200 &\texttwelveudash & \texttwelveudash & 631.2 & \texttwelveudash & 30 & $0.0931$   \\ \cline{2-9}
& Green & 4200 & \texttwelveudash & \texttwelveudash & 616.9 & \texttwelveudash & 30 & $0.0975$ \\ \hline

\end{tabular}
 \label{tbl:a}
\end{table}
\end{center}

Normally, the OPL spectra have Lorentzian line-shape when a gold nanoparticle is excited by a laser field blue-detuned to the peak of scattering  \cite{NL12p4385, ChemPnt, PulsePL1, SizeDependence1}. However, we find that the anti-Stokes OPL emission spectra is not a Lorentzian profile when a red-detuned excitation laser is applied.
In contrast to the scattering, a cw laser field at $632.8$ \nano\meter~ was applied in our experiment to excite the OPL. Because the intensity, $I_\text{in}$, and wavelength, $\lambda_\text{in}=632.8$ \nano\meter, of the excitation laser in experiments are fixed, we fit all OPL spectra using the same function for FESD, $\rho_\text{e}(\lambda) = N_\text{FE}/(1+e^{\tau_T(\lambda_\mu/\lambda -1)})$ with $\lambda_\mu = 633$ \nano\meter~ and $\tau_T = 65$. For the Stokes component, the FESD distribution is flat approximately and the OPL spectral profile is dominated by the LSPR mode. As a result, the Stokes spectrum has Lorentzian profiles. Remarkably, the anti-Stokes components are dominated by $\rho_\text{e}(\lambda)$, which is the Fermi-Dirac distribution multiplied by the total number of free electrons $N_\text{FE}$. our model is in good agreement with our experimental observation. Obviously, it reveals that the free-electron state density in thermal equilibrium crucially modifies the anti-Stokes components of the PL emission from a gold nanorod.

The constant factors $A^\prime\kappa^2$ and $A\kappa^3$ for the scattering and the OPL spectra are close when we change the refractive index in \emph{situ}. Their relative difference is about $30\%$ and can attribute to the change of the polarization of the LSPR mode and the absorption of environment. The cases for the reshaped nanorods are essentially different. The wavefunction $|H_\text{COFE}\rangle$ and the polarization $\vec{d}$ of the LSPR mode are crucially dependent on the shape of nanoparticles, whereas the input laser field is linear-polarized and has a fixed mode profile $|H_\text{photon}\rangle$. As a result, the factors $A\kappa^3$ in the PL spectra changes considerably during reshaping.

\begin{center}
\begin{table}[H]
\centering
\caption{Parameters for scattering and PL spectra in Figs.~\ref{fig:PL} (c) and (d)}
\begin{tabular}{l|c|c|c|c|c|c|c|r}
\hline \hline
 & Line \# & $AN_{FE}\cos^2\beta\cos{\theta}^2\eta_c(\omega_{in})$ & $I_{in}$ [counts]& $\lambda_{in}$ [\nano\meter] & $\lambda_{c}$ [\nano\meter] & $\lambda_{in} - \lambda_{c}$ [\nano\meter] & $\delta\lambda_c$ [\nano\meter] & $\kappa$ [\electronvolt] \\ \hline
 \multirow{2}{*}{PL.(c)} & Green & $1$ & $1.38 \times 10^4$ & $632.8$ & $624$ & $8.8$ & $20.5$ & $0.065$   \\ \cline{2-9}
 & red & $1.5$  & $1.38 \times 10^4$ & $632.8$ & $660$ & $-27.2$ & $22$ & $0.0625$  \\ \hline \hline
 \multirow{2}{*}{PL.(d)} & blue & $1.06$ & $1.38 \times 10^4$  & 632.8 & 640 & -7.2 & 21 & $0.0634$  \\ \cline{2-9}
 & red & $0.58$ & $1.38 \times 10^4$ & 632.8 & 627 & 5.8 & 21.5 & $0.0677$  \\ \cline{2-9}
 &  Green & $0.77$ & $1.38 \times 10^4$ & 632.8 & 615.4 & 17.4 & 22 & $0.0719$ \\ \hline
\end{tabular}
 \label{tbl:b}
\end{table}
\end{center}

The comparison between our model and experimental observation shows that our model provides a unified understanding of the experimental observation for both the anti-Stokes and Stokes PL emission. It reveals that the quantum occupation of free-electron state $\rho_\text{e}(\omega)$ in the thermal equilibrium essentially modifies the anti-Stokes components of the PL emission from a single gold nanorod. We note that the anti-Stokes radiation has been reported as a result of the thermal radiation \cite{thermal1} or the thermal population of the electron-hole excitations during Raman scattering \cite{thermal2}. However, it is improper to attribute the OPL emission observed in our experiment to the thermal radiation simply. The thermal radiation from a black body \cite{thermal1} is unpolarized and must be polarization-independent. On the contrary, the OPL intensity in our experiment is crucially dependent on the polarization of the exciting laser and the collection polarization of the detector \cite{OurExperiment}. Moreover, the polarization dependence of the anti-Stokes emission is strongly correlated to that of the Stokes emission. It implies that the anti-Stokes and the Stokes emission in our experiment share a common mechanism fundamentally different from the thermal radiation. Besides, the model based on the thermal distribution of the electron-hole excitation only explain the anti-Stokes emission with a large Raman shift \cite{thermal2}. The discrepancy between the model and the experimental observation results to the deviation at the Stokes emission and even at the anti-Stokes near the excitation frequency which can be one order in amplitude \cite{thermal2}. A weak anti-Stokes PL has been observed in the OPL of a single gold nanoparticle by Beversluis et al. \cite{PRB68p115433}, but an explanation is lack. Here we present a self-consistent and unified model in good agreement with both the anti-Stokes and Stokes components in the PL from a single gold nanorold.

\begin{figure}
 \centering
 \includegraphics[width=0.45\linewidth]{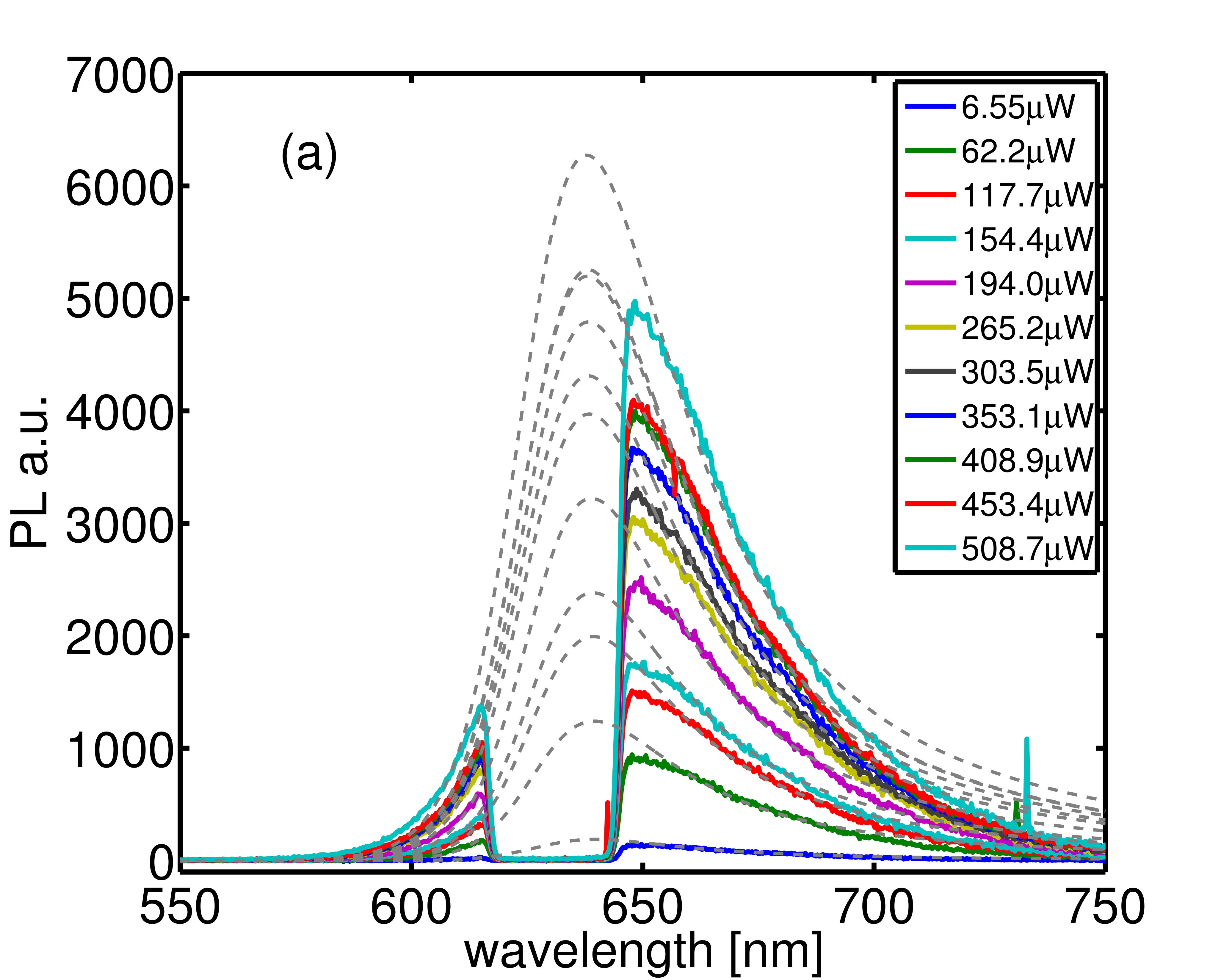}
 \includegraphics[width=0.52\linewidth]{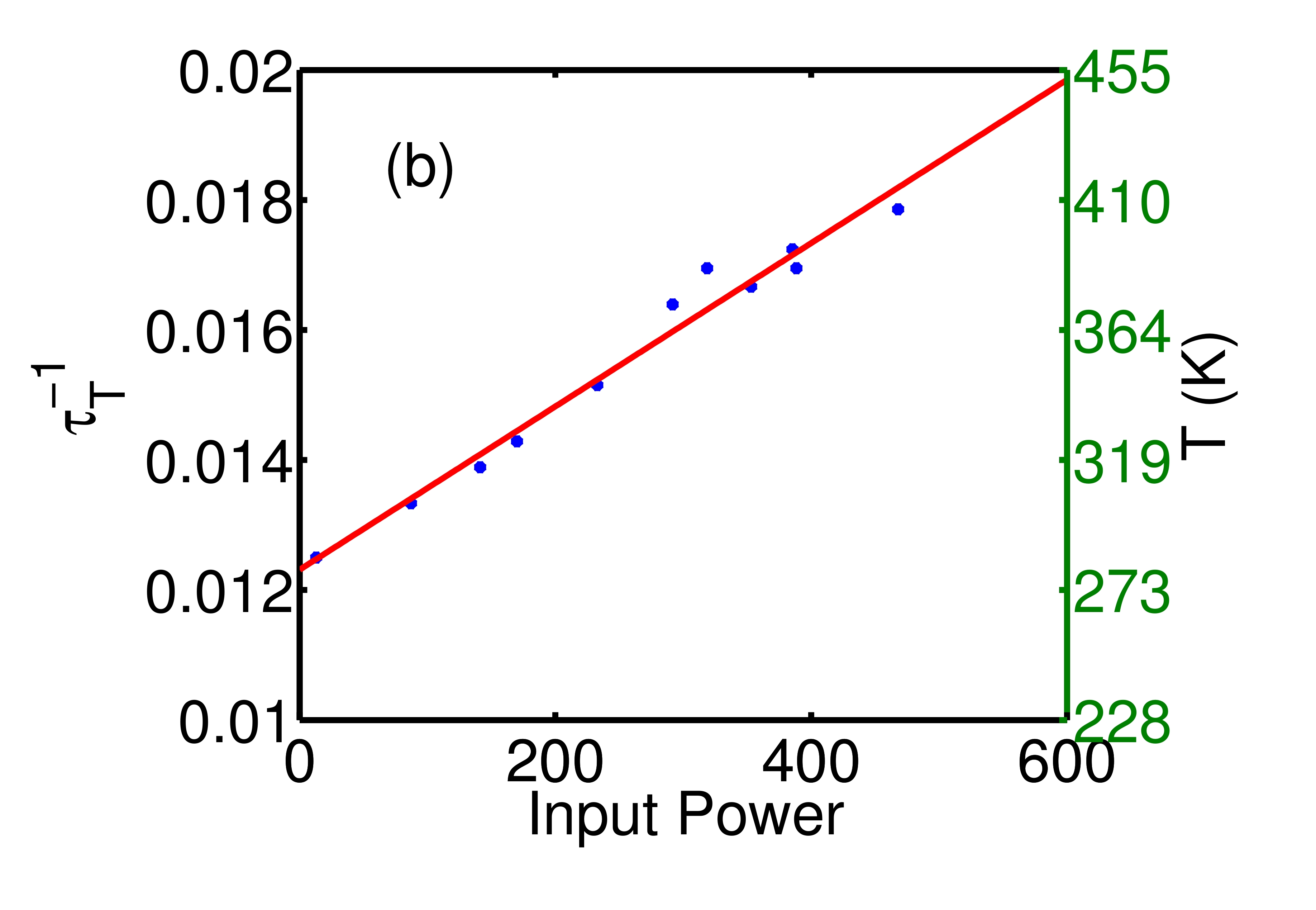}\\
 \caption{(Color online) Dependence of temperature on the input power. (a) Fitting the PL spectra from a single gold nanorod excited by a cw laser with different power $P_\text{in}$. To fit the data, we have scaled the input power by a factor $\sim 1$. (b) Dependence of the parameter $\tau_T$ and the local temperature $T$ in the FESD $\rho_\text{e}(\omega)$ as a function of the input power. $\lambda_\mu=633$ \nano\meter.}\label{fig:tau}
\end{figure}

An interesting parameter in our model is $\tau_T$ corresponding to the nanoscale-localized temperature. We studied the dependence of $\tau_T$ on the input power by fixing the frequency of cw excitation. We increased the power of the laser by two orders from $\sim 7$ \micro\watt~ to $\sim 500$ \micro\watt. By fitting the experimental data shown in Fig. \ref{fig:tau} (a), we find a relation $\tau_T^{-1}$ corresponding to the nanoscale-localized temperature $T$ increases linearly as $P_{in}$ increases (see Fig. \ref{fig:tau} (b)), where $P_{in}$ is the power of the excitation laser. Note that the maximal temperature of our samples is below $410$ \kelvin. Such low temperature can not support an effective thermal luminescence at the visible light frequency. The thermal radiation even at $500$ \kelvin~has a peak around $0.13$ \electronvolt~ and the luminescence with energy larger than $0.65$ \electronvolt~ is vanishing. In contrast, our experimental measurements show the OPL spectra with peaks around $640$ \nano\meter~ ($\sim 1.9$ \electronvolt) with all energy larger than $800$ \nano\meter~ ($\sim 1.55$ \electronvolt). The quantitative agreement betwen the experiment measurement and our theory provides a way to measure the nanoscale localized temperature. As shown in  Fig. \ref{fig:tau} (b), the temperature of the gold nanorod is about $275$ \kelvin when $P_{in}=6.55$ \micro\watt. It is evidenced that the gold nanorod is not heated by the weak cw laser if $P_{in}\lesssim 6.5$ \micro\watt. As a result, the temperature of the gold nanorod we measure reflects that of nanoscale localized environment. On the other hand, we can control the local temperature of environment by heating a gold nanorod because the temperature of nanoparticle is linearly dependent on the input laser power. The  Fermi-Dirac redistribution of FESD in our model indicates a physic mechanism fundamentally different from the thermal excitation suggested by Huang et al. \cite{LightEmission1}. This process has been accepted to explain the frequency up-conversion in luminescence from single noble metal nanoparticles \cite{HotE}. 

\subsection{Theoretical prediction of quantum yield}

The quantum yield of the OPL from noble metal nanoparticles is a few orders higher in amplitude in comparison with bulk film \cite{HighQY1,PulsePL1,SizeDependence1,NL12p4385} and has shape-dependence \cite{NL12p4385,SizeDependence1,SizeDependence2}. However, ``no explanation was proposed for the larger QY of nanorods and its shape dependence.'' and the origin of the saturation of the PL QY is unknown \cite{NL12p4385}. Our model provides a understanding to the origin of the enhanced QY and the saturation phenomenon.

Now we compare the QYs between a single nanorod and a bulk film. We consider the bulk film as a long nanorod with a very small $\omega_\text{c}(L) \ll \omega_\text{in}$. Therefore, at a given detection frequency $\omega$, the ratio of the PL intensity between the single nanorod and the bulk film is roughly $\omega_\text{in}^2\omega^2/\left[(\omega_\text{in}-\omega_\text{c}(L))^2 + \kappa^2\right] \left[(\omega-\omega_c(L))^2 + \kappa^2\right]$ ($\omega_\text{in},\omega \gg \kappa$). When $\omega_\text{in}\sim\omega_\text{c}(L)$ and $\omega\sim\omega_\text{c}(L)$, the ratio is largest and can be the order of $Q^4$ where $Q=\omega_\text{c}(L)/\kappa$ is the quality factor of the scattering spectrum. A mediate quality factor $Q=20$ already yields a ratio of $1.6\times 10^5$. According to our model, the high PL QY of noble metal nanorods origins from the enhancement of the excitation and emission due to the resonance plasmonic resonator similar to the case that the optical cavity greatly increases the localized optical field.

\begin{figure}
 \centering
 \includegraphics[width=0.42\linewidth]{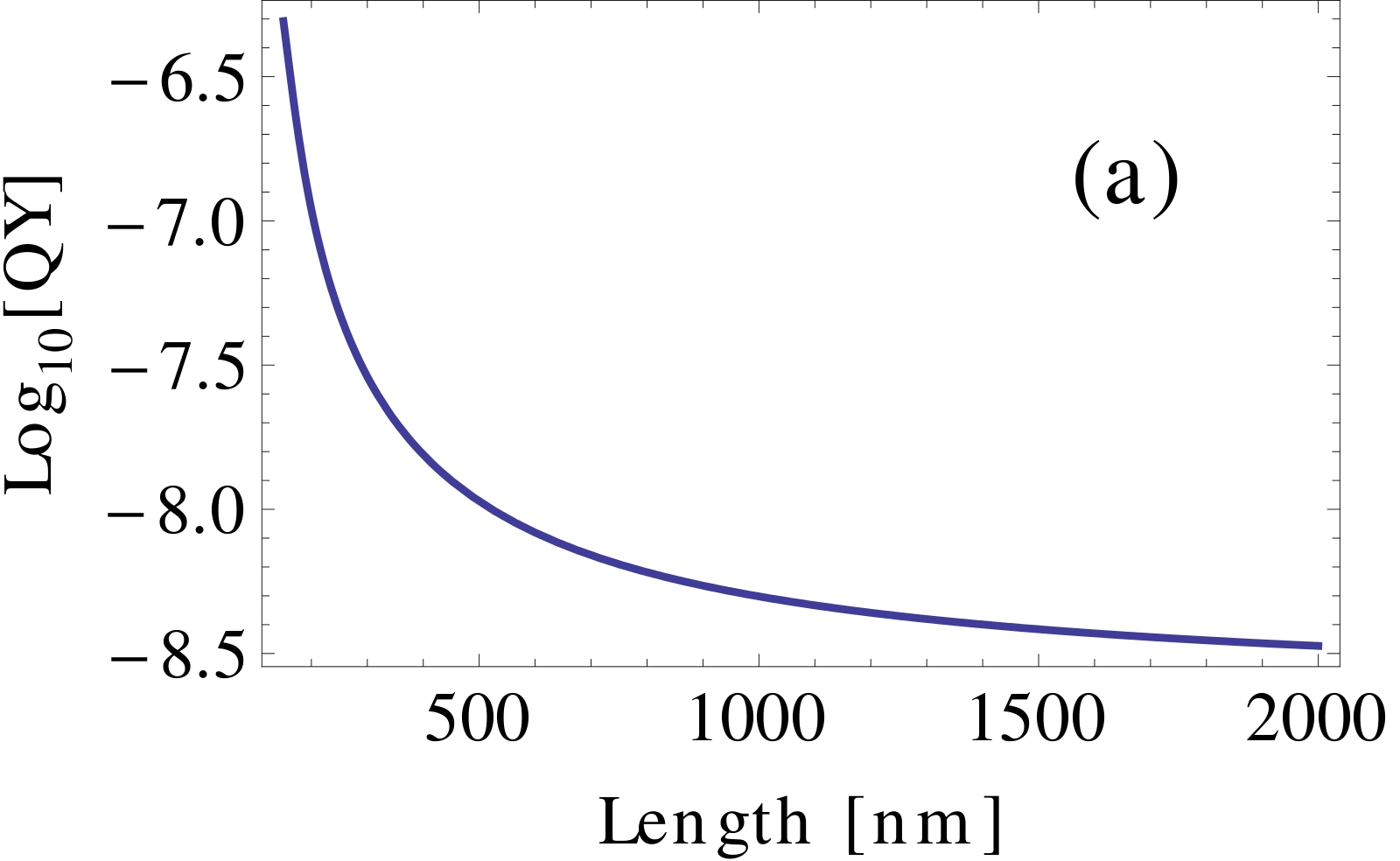}
 \includegraphics[width=0.42\linewidth]{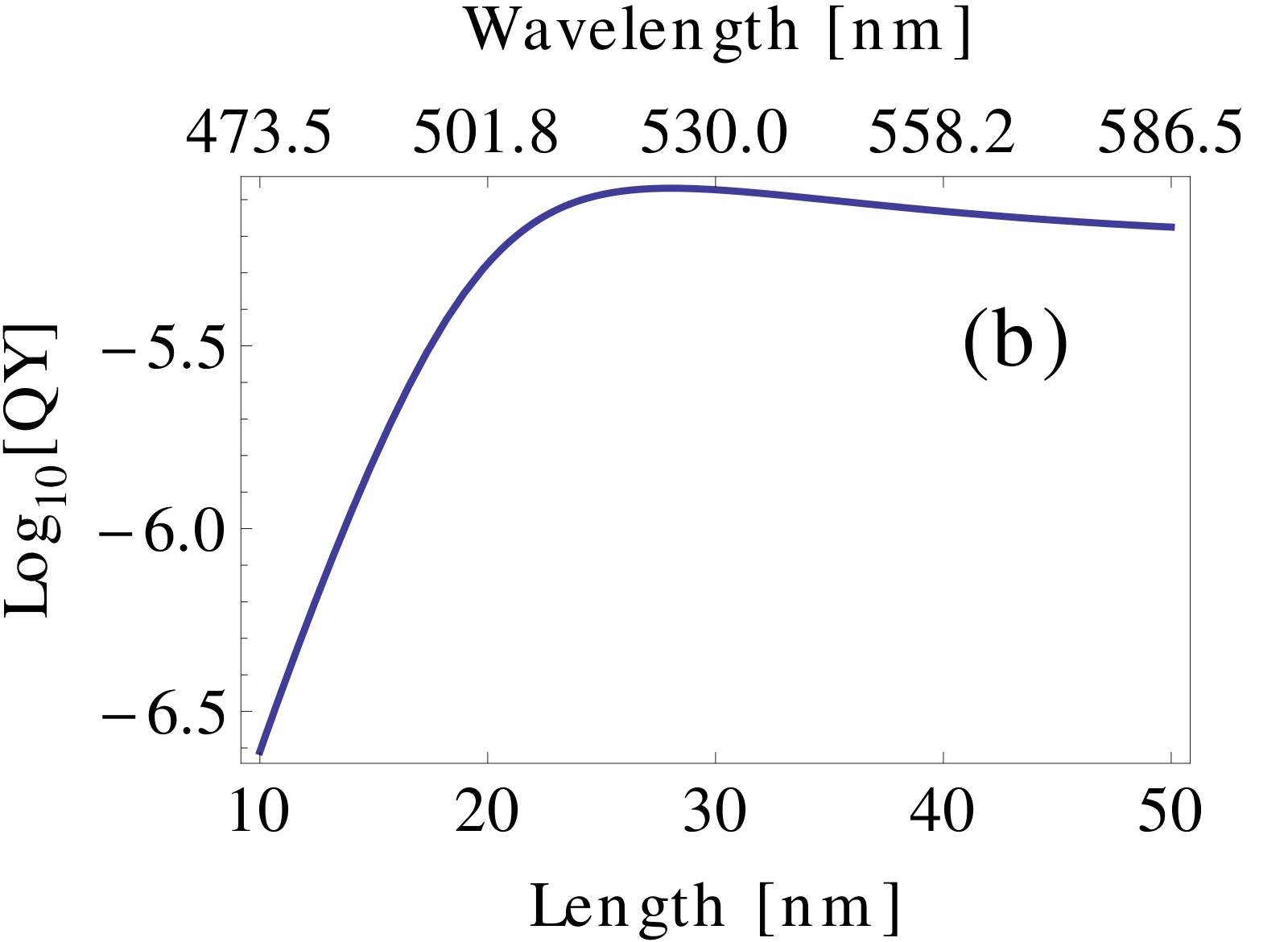}\\
 \caption{(Color online) Quantum yield of the PL emission. (a) a bulk material, (b) a single gold nanorod. Other parameters are $\omega_\text{in}=2.474$ \electronvolt, $\tau_T=65$. For simplicity, we fix $\kappa=0.123$ \electronvolt. To provide a quality reproduction of the measurement \cite{NL12p4385}, we simply assume a linear relation $\lambda_\text{c}=530 \nano\meter + 2.82(L-30\nano\meter)$ as an example \cite{SizeDependence1}. Note that this relation depends on the shape of the samples.}\label{fig:BulkNano}
\end{figure}

The QY as a sum of the OPL spectrum is dependent on the length of nanorods. In Fig. \ref{fig:BulkNano}, we numerically calculate the QY for single nanorods and a bulk film. Clearly, the QY of a bulk film can be as low as $10^{-8.5}$ and that of a single nanorod can be two to three orders higher. More importantly, our model correctly predicts the observed saturation of the QY of the OPL from a single gold nanorod \cite{NL12p4385} even without taking into account the free path effect of the electrons \cite{MFEP}. We find that the QY increases first when $L$ is small and then becomes saturate at $\sim L_\text{s}=25$ \nano\meter~. Note that the saturation length $L_s$ is dependent on the specific function $\lambda_\text{c}(L)$. Because we are lack of the knowledge of $\lambda_\text{c}(L)$ in experiment \cite{NL12p4385}, our model can only provide a quality prediction.

\begin{figure}
 \centering
 \includegraphics[width=0.45\linewidth]{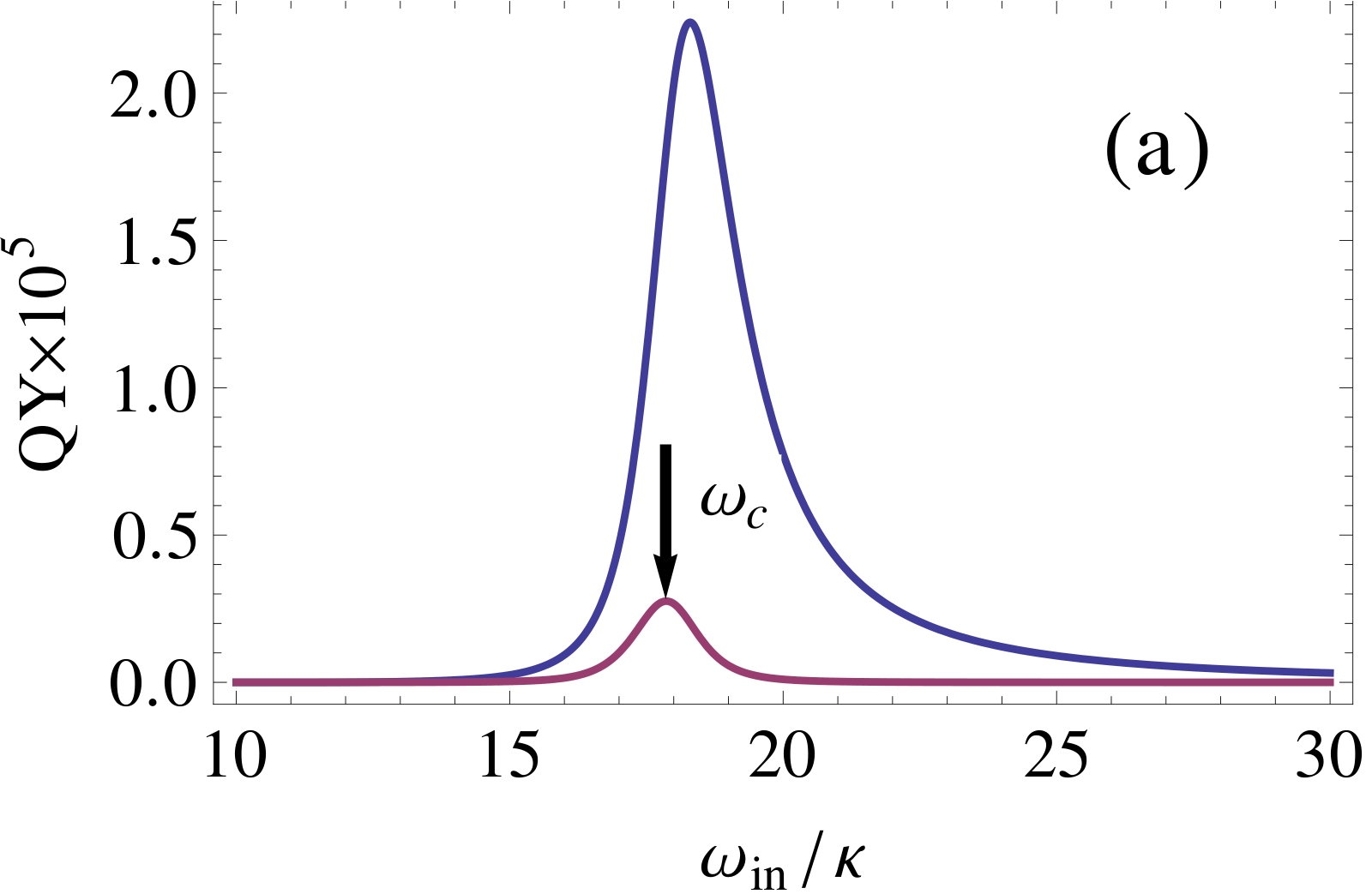}
 \includegraphics[width=0.45\linewidth]{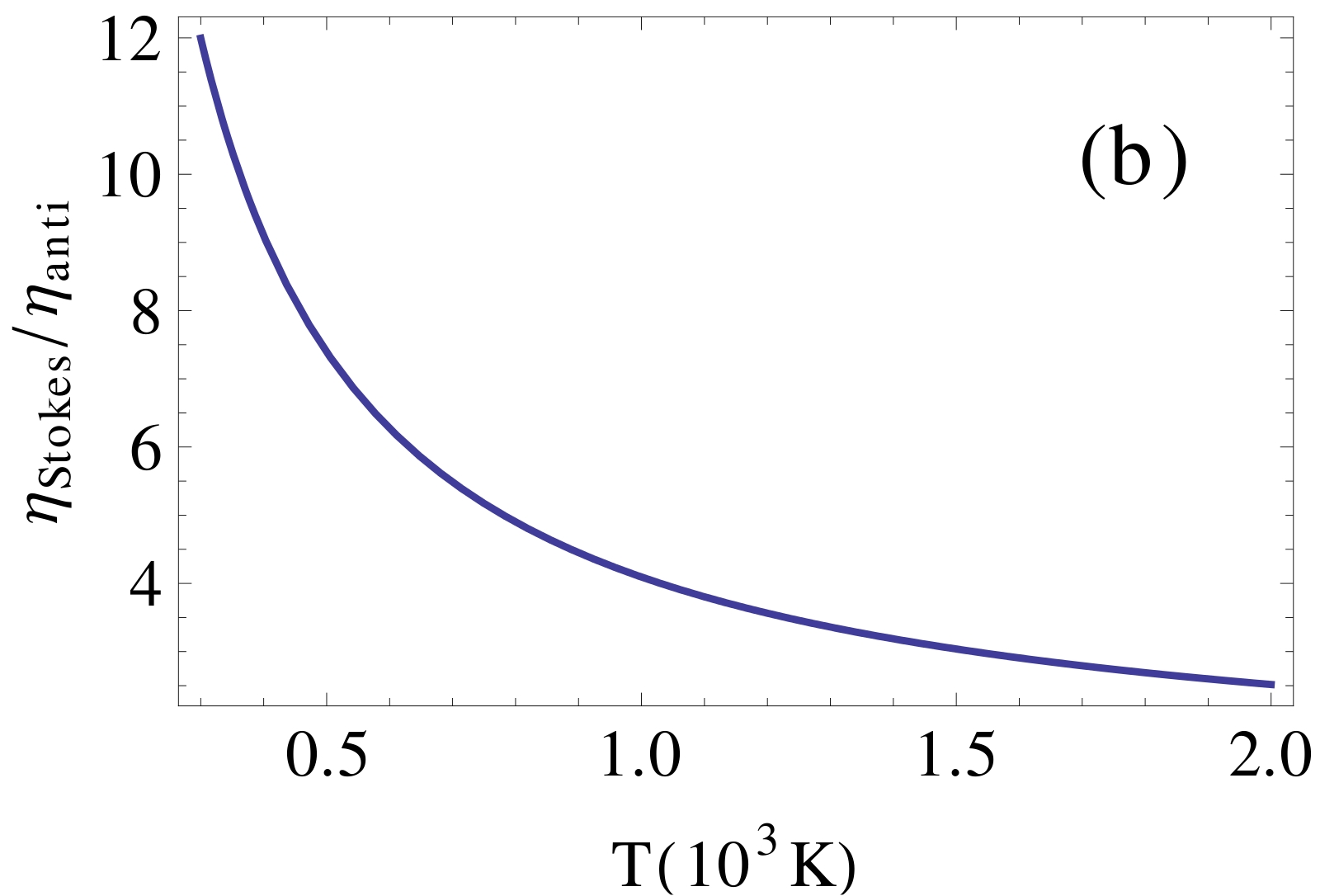}\\
 \caption{(Color online) Comparison of the anti-Stokes and Stokes emission. (a) Quantum yield of the anti-Stokes and Stokes components and the PL emission from a single gold nanorod as a function of the excitation frequency for $\tau_T=65$. (b) Ratio of quantum yields between the anti-Stokes and Stokes for $\hbar\omega_\text{in}=2.474$ \electronvolt. Other parameters are $\hbar\omega_\text{c}=2.22$ \electronvolt, $\kappa\approx 0.123$ \electronvolt.}\label{fig:TwoComp}
\end{figure}

According to Eqs. \ref{eq:QYS} and \ref{eq:QYAS} both the anti-Stokes and Stokes components of the PL QY are dependent on the excitation frequency $\omega_\text{in}$. It can be seen from Fig. \ref{fig:TwoComp} (a) that both components increase first and then decrease as the excitation frequency increases. It means that both the anti-Stokes and Stokes intensities decreases as the detuning increases. When the excitation matches the resonance frequency of the plasmonic resonator $\omega_\text{c}$, both the anti-Stokes and Stokes reach the peak. However, the anti-Stokes emission is always smaller than the Stokes one. The largest ratio between two components is about $0.22$ at $\omega_\text{in}=0.95\omega_c$ for $\tau_T=65$ corresponding to $T=441$ \kelvin. As shown in Fig. \ref{fig:TwoComp} (b), increasing the temperature of the single gold nanorod can increase the efficiency of the anti-Stokes emission because the distribution of FESD becomes flatter and flatter as the temperature increasing. For example, the ratio decreases from $7.4$ to $2.5$ as the temperature increases from $500~ \kelvin$ to $2000~ \kelvin$. Note that the incoherent thermal radiation of photon energy larger than $\omega_\text{in}$ is still negligible even at $T=2000~ \kelvin$ according to the Planck distribution. The coherent radiation due to the COFE is dominant.

\subsection{Two-photon luminescence}
\begin{figure}
 \centering
 \includegraphics[width=0.42\linewidth]{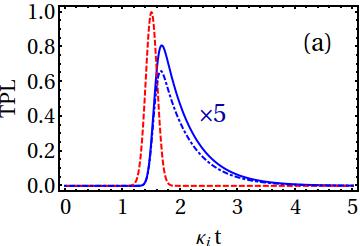}
 \includegraphics[width=0.45\linewidth]{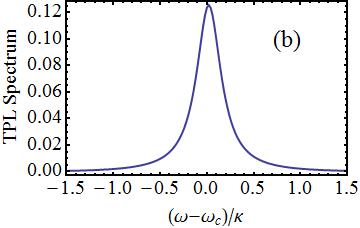}\\
 \caption{(Color online) Two-photon luminescence from a single gold nanorod excited by a ultrashort laser pulse. (a) Temporal TPL intensity. Blue solid (dotted-dashed) line is the TPL intensity (increased by five fold) for $\Delta = \omega_\text{c} - 2\omega_\text{in}=0 (5\kappa_\text{i})$; dashed red line for a Gaussian excitation laser pulse with a duration $\tau_\text{in}=0.1\kappa_i$. (b) TPL spectrum calculated as Fourier transform of TPL intensity (blue solid line). $\kappa=\kappa_\text{i} + \kappa_\text{ex}$ and $\kappa_\text{i}= \kappa_\text{ex}$. }\label{fig:TPL}
\end{figure}

Now we use our model to study the time dynamics and spectral feature of the TPL from a single gold nanorod illuminated by a ultrashort laser pulse. In an ideal case such that the laser pulse can be considered as an instant excitation, the TPL intensity exponentially decays and its spectrum is a simple Lorentzian profile.
In a real case, the duration of laser pulse is finite. We need solve Eq. \ref{eq:TPLt} to calculate the time dynamics of the intensity of TPL and the spectrum. As an example shown in Fig. \ref{fig:TPL}, we excite TPL with a Gaussian ultrashort laser pulse with a duration $\tau_\text{in}$ much shorter than the ringdown time $\kappa^{-1}$ of the single gold nanorod. It can be seen that the TPL intensity is first excited by the laser pulse to a high level and then decays exponentially with a rate of $\kappa$. A finite detuning $\omega_c -2\omega_\text{in}$ between the resonator mode and the two-photon energy leads to a decrease in the TPL intensity. Unlike the phenomenological explanation, our model based on a microscopic plasmonic resonator concept predicts the exponential decay behavior of TPL. This prediction agrees well with experimental observation \cite{thermal2, PulsePL4, TPL2}. As a result of an exponential decay, the spectrum of TPL is approximately a Lorentzian profile resembling the scattering spectrum. 

We are aware of that the PL spectrum from gold nanorods are normally complex in experiments and may has more than one resonance peak. However, considering only the longitudinal mode, our theory presents a simple but basic concept for understanding the main features of the photoluminescence from single gold nanorods.

\FloatBarrier
\section{CONCLUSION}

In summary, we have experimentally observed the one-photon luminescence emission from a single gold nanorod modeled as a plasmonic resonator. Based on experimental observations, we have presented a unified theoretical model to explain both the anti-Stokes and Stokes spectral components although their profiles are substantially different. Our work reveals that the extraordinary enhancement of the OPL from a single gold nanorod origins from the enhanced interaction between the laser and the collective oscillation of free-electrons as a resonance mode in a plasmonic nanoresonator. The Stokes OPL has Lorentzian profile and origins mainly from the resonant emission of the plasmonic nanoresonator, while the anti-Stokes OPL is the result of the radiation from the plasmonic nanoresonator modulated strongly by the redistributed free-electron state density. The anti-Stokes OPL emission process is determined by both the cavity resonance and electron distribution. Our model also provides a unified explanation for the polarization and size dependence of the OPL from a single gold nanorod. In particular, it predicts the enhancement and the saturation of the QY. Furthermore, the dynamics and spectral profile of the TPL are also correctly explained by our model. Interestingly, both experimental and theoretical results demonstrate a new way to sense or control the temperature of environment localized in a nanoscale.
\newline

\noindent \textbf{Acknowledgement}\\
This work was supported by the National Key Basic Research Program of China (grant No. 2013CB328703) and the National Natural Science Foundation of China (grant Nos. 11204080, 11374026, 61422502, 61521004, 91221304). K. X. also thanks for the support by the Australian Research Council Centre of Excellence for Engineered Quantum Systems (EQuS), project number, CE110001013. \newline

%


\end{document}